\documentclass[aps,10pt,prd,preprintnumbers,amsmath,amssymb,nofootinbib,superscriptaddress,a4paper,showpacs]{revtex4-1}
\pdfoutput=1
\usepackage[T1]{fontenc}

\usepackage[english]{babel}
\usepackage[utf8]{inputenc}
\usepackage{graphicx}
\usepackage{color}
\usepackage{amsfonts,amsthm}
\usepackage{bm,bbm}
\usepackage{xcolor}
\usepackage{mathrsfs}
\usepackage{amsmath}

\newcommand{\cF}{{\cal F}}
\newcommand{\cL}{{\cal L}}
\newcommand{\cM}{{\cal M}}
\newcommand{\cN}{{\cal N}}
\newcommand{\cO}{{\cal O}}
\newcommand{\cP}{{\cal P}}

\newcommand{\cV}{{\cal V}}
\newcommand{\cT}{{\cal T}}
\newcommand{\cW}{{\cal W}}
\newcommand{\cY}{{\cal Y}}

\newcommand{\lambdab}{{\overline{\lambda}}}
\newcommand{\phib}{{\overline{\phi}}}
\newcommand{\Tr}{{\rm Tr\;}}
\newcommand{\hf}{\frac{1}{2}}
\newcommand{\qtr}{\frac{1}{4}}
\newcommand{\rtwo}{\sqrt{2}}
\newcommand{\bn}{{\bf n}}
\newcommand{\tQ}{\widetilde{Q}}

\def\nn{\nonumber}
\def\bec{\begin{center}}
\def\eec{\end{center}}
\def\beq{\begin{equation}}
\def\eeq{\end{equation}}
\def\bea{\begin{eqnarray}}
\def\eea{\end{eqnarray}}

\begin{document}

\title{Lattice Formulation of $\cN = 2^*$ Yang-Mills}

\author{Anosh Joseph}

\affiliation{International Centre for Theoretical Sciences (ICTS-TIFR), \\
				Tata Institute of Fundamental Research, \\
				Shivakote, Bangalore 560089 INDIA} 

\date{\today}

%%%%%%%%%%%%%%%%
\begin{abstract}
%%%%%%%%%%%%%%%% 

We formulate ${\cal N} = 2^*$ supersymmetric Yang-Mills theory on a Euclidean spacetime lattice using the method of topological twisting. The lattice formulation preserves one scalar supersymmetry charge at finite lattice spacing. The lattice theory is also local, gauge invariant and free from doublers. We can use the lattice formulation of ${\cal N} = 2^*$ supersymmetric Yang-Mills to study finite temperature nonperturbative sectors of the theory and thus validate the gauge-gravity duality conjecture in a nonconformal theory.

%%%%%%%%%%%%%%
\end{abstract}
%%%%%%%%%%%%%%


\maketitle

%%%%%%%%%%%%%%%%%%%%%%%%%%%%%%%%%%%%%%%%%%%%%%%%%%%%%%%
\section{Introduction}
\label{sec:intro}
%%%%%%%%%%%%%%%%%%%%%%%%%%%%%%%%%%%%%%%%%%%%%%%%%%%%%%%

Supersymmetric quantum field theories are interesting classes of theories by themselves. They can also be used to construct many phenomenologically relevant models such as the minimal supersymmetric standard model. Supersymmetric quantum field theories exhibit many interesting features when they are strongly coupled. It is in general difficult to study analytically the strong coupling regimes of supersymmetric quantum field theories. If we could formulate such theories on a spacetime lattice, in a consistent manner, we would have a first principles definition of the theory that can be used to study their nonperturbative sectors. Certain classes of supersymmetric field theories can be formulated on a spacetime lattice by preserving a subset of the supersymmetry charges. These approaches are based on the methods of topological twisting \cite{Witten:1988ze} and orbifolding \cite{ArkaniHamed:2001ca}, and they can be used to formulate lattice theories with extended supersymmetries. 

Supersymmetric lattices have been constructed for several classes of theories in various spacetime dimensions \cite{Sugino:2003yb, Catterall:2005fd, Unsal:2006qp, Catterall:2007kn, Catterall:2009it, Catterall:2011pd, Joseph:2011xy, Catterall:2011aa, Catterall:2013roa, Joseph:2013bra, Joseph:2014bwa, Catterall:2014vka, Joseph:2014wxa, Joseph:2015xwa, Joseph:2016tlc, Joseph:2016cdq}, including the well-known $\cN = 4$ supersymmetric Yang-Mills (SYM) theory in four spacetime dimensions \cite{Sugino:2003yb, Catterall:2005fd, Kaplan:2005ta}.

In this work we provide the lattice construction of a very interesting theory, which is known as the four-dimensional $\cN = 2^*$ supersymmetric Yang-Mills theory \cite{Polchinski:2000uf}. This nonconformal field theory is obtained by giving mass to the hypermultiplet of four-dimensional $\cN = 4$ SYM theory. The $\cN = 2^*$ SYM theory also takes part in the AdS/CFT correspondence. Its gravitational dual has been constructed by Pilch and Warner \cite{Pilch:2000ue}.

In the recent past, supersymmetric lattice constructions have been used to test and validate the gauge-gravity duality conjecture in various dimensions \cite{Catterall:2009xn, Catterall:2010fx, Catterall:2010ya, Kadoh:2015mka, Filev:2015hia, Asano:2016xsf, Berkowitz:2016jlq, Asano:2016kxo, Catterall:2017lub}. Those lattice studies gave consistent results with other approaches \cite{Hanada:2007ti, Hanada:2008gy, Hanada:2008ez}. See also Refs. \cite{Catterall:2011cea, Schaich:2014pda} for computer codes developed for simulating SYM theories with 4 and 16 supercharges in spacetime dimensions $d \leq 4$.     

We use the method of topological twisting to construct $\cN = 2^*$ SYM on a Euclidean spacetime lattice. The continuum twisted $\cN = 2^*$ SYM theory is obtained by introducing mass deformation terms to the Vafa-Witten twisted $\cN = 4$ SYM theory \cite{Vafa:1994tf}. Once we have a twisted version of $\cN = 2^*$ SYM theory in the continuum it is straightforward to implement the theory on the lattice. We use the discretization prescription provided by Sugino \cite{Sugino:2003yb}. The lattice formulation preserves one supersymmetry charge at finite lattice spacing. The lattice construction is also local, gauge invariant and free from the problem of fermion doublers. The mass deformation, however, introduces terms in the action that are not twisted Lorentz invariant. This is related to the fact that the R-symmetry group of $\cN = 2^*$ SYM theory, $SU(2) \times U(1)$, is smaller than its Euclidean Lorentz rotation symmetry group, $SO(4)$. The presence of twisted Lorentz noninvariant terms does not lead to any inconsistencies in the theory since the twisted theory is still Lorentz invariant due to the fact that twisting is just an exotic change of variables in flat Euclidean spacetime.         

One could use the lattice construction of $\cN = 2^*$ SYM presented here to explore the nonperturbative sectors of the theory, including its thermodynamic properties, and compare with the existing results from the dual gravitational theory. 

The paper is organized as follows. In Sec. \ref{sec:4d_n4_sym} we briefly review the four-dimensional $\cN = 4$ SYM in flat Euclidean spacetime. We also review how the mass deformation of $\cN = 2$ hypermultiplet gives rise to $\cN = 2^*$ SYM. In Sec. \ref{sec:Vafa-Witten-twist}, we review the Vafa-Witten twist of $\cN = 4$ SYM, which is relevant for the lattice formulation of the $\cN = 2^*$ SYM. Although there exist two more inequivalent twists of $\cN=4$ SYM, we believe that they are not suitable for the construction of lattice regularized $\cN = 2^*$ SYM. After these warm-up and review sections, we move on to constructing the twisted version of $\cN=2^*$ SYM in the continuum in Sec. \ref{sec:twisted-n2*-sym}. This is the first time such a continuum twisted formulation of $\cN=2^*$ SYM is presented according to our knowledge. In Sec. \ref{sec:lattice}, we introduce the lattice formulation of this theory. It is convenient to write down the twisted theory in a form known as the balanced topological field theory form (BTFT) before lattice regularization. After expressing the theory in a BTFT form we move on to the details of the lattice formulation. We use the lattice discretization prescription given by Sugino. The lattice action can be expressed as a twisted scalar supersymmetry variation of a gauge fermion. The lattice construction is gauge invariant, local and doubler free and preserves one supersymmetry charge at finite lattice spacing. We end with conclusions and future directions in Sec. \ref{sec:conclusions-future}.    

%%%%%%%%%%%%%%%%%%%%%%%%%%%%%%%%%%%%%%%%%%%%%%%%%%%%%%%
\section{$\cN = 4$ SYM and Mass Deformation to $\cN = 2^*$ SYM}
\label{sec:4d_n4_sym}
%%%%%%%%%%%%%%%%%%%%%%%%%%%%%%%%%%%%%%%%%%%%%%%%%%%%%%%

We briefly review the $\cN = 4$ SYM on flat ${\mathbb R}^4$. In the language of $\cN = 1$ superfields, $\cN = 4$ SYM theory contains one vector multiplet and three adjoint chiral multiplets. We denote them as superfields $V$ and $\Phi_s$, with $s = 1, 2, 3$.

The physical component fields of the superfields are
\beq
\begin{aligned}
V &\longrightarrow A_\mu,~\lambda_{4\alpha},~\lambdab^4_{\phantom{4}\dot{\alpha}}~, \\
\Phi_s, \Phi^{\dagger s} &\longrightarrow \phi_s,~\lambda_{s\alpha},~\phi^{\dagger s},~\lambdab^s_{\phantom{s}\dot{\alpha}}~.
\end{aligned}
\eeq

The theory has global symmetry group 
\beq
SU(2)_L \times SU(2)_R \times SU(4)~,
\eeq
where $SU(2)_L \times SU(2)_R \simeq SO(4)$ is the Euclidean Lorentz rotation group and $SU(4) \simeq SO(6)$ denotes the R-symmetry group. 

The gauge field $A_\mu$ is a scalar under $SU(4)$. The gauginos $\lambda_{s\alpha},~\lambdab^s_{\phantom{s}\dot{\alpha}}$ and the six scalars $\phi_s,~\phi^{\dagger s}$ transform as ${\mathbf 4} \oplus \overline{{\mathbf 4}}$ and ${\mathbf 6}$, respectively under $SU(4)$ internal rotation symmetry. The scalars can be packaged into an antisymmetric and self-conjugate tensor $\phi_{uv}$ with $u, v = 1, 2, 3,4$ representing the indices of the fundamental representation of $SU(4)$. In this notation the gauginos of vector and chiral multiplets can be combined: $\lambda_{u\alpha},~\lambdab^u_{\phantom{u}\dot{\alpha}}$. All fields of the theory take values in the adjoint representation of the gauge group. Here we take the gauge group to be $SU(N)$. We use the anti-Hermitian basis for the generators of the gauge group, with the normalization $\Tr (T_a T_b) = - \delta_{ab}$.

We can combine the superfield $V$ and one of the adjoint chiral superfields, say, $\Phi_3$ to form an $\cN = 2$ vector multiplet. The chiral superfields $\Phi_1$ and $\Phi_2$ can be combined to form an $\cN = 2$ hypermultiplet. The $\cN = 2^*$ SYM theory is a one-parameter (real) mass deformation of $\cN = 4$ SYM obtained by giving mass to the fields of the $\cN = 2$ hypermultiplet. The mass terms softly break supersymmetry from $\cN = 4$ to $\cN = 2$. The $\cN = 2^*$ SYM theory has a fixed point in the far UV, which is the conformal $\cN = 4$ SYM theory. The mass deformation is relevant and it induces running in the coupling, so that the theory becomes pure $\cN = 2$ SYM in the deep IR.

On flat ${\mathbb R}^4$, the mass deformation takes the following form in terms of the component fields \cite{Labastida:1997xk, Buchel:2012gw},
\bea
\label{eq:n2-star-untwisted-mass}
S_m &=& \frac{1}{g^2} \int d^4x~ \Tr  \Big( - m \lambda_1^{\phantom{1}\alpha} \lambda_{2\alpha} - m \lambdab^1_{\phantom{1}\dot{\alpha}} \lambdab^{2\dot{\alpha}} + m^2 \phi_1^{\phantom{\dagger}} \phi_1^\dagger + m^2 \phi_2^{\phantom{\dagger}} \phi_2^\dagger \nn \\
&& \quad \quad \quad \quad \quad \quad - \rtwo m \phi_3^{\phantom{\dagger}}[\phi_1^{\phantom{\dagger}}, \phi_1^\dagger] - \rtwo m \phi_3 [\phi_2^{\phantom{\dagger}}, \phi_2^\dagger] \nn \\
&& \quad \quad \quad \quad \quad \quad - \rtwo m \phi_3^\dagger[\phi_1^{\phantom{\dagger}}, \phi_1^\dagger] - \rtwo m \phi_3^\dagger [\phi_2^{\phantom{\dagger}}, \phi_2^\dagger] \Big)~,
\eea
where $m$ is the mass parameter and $g$ is the coupling constant of the theory. The deformation gives conventional mass terms for two Weyl fermions and two complex scalars and also trilinear couplings between the $\cN = 2$ hypermultiplet scalars and the vector multiplet scalar $\phi_3$. 

Motivated by the supergravity dual geometry of $\cN = 2^*$ Yang-Mills theory,\footnote{In Appendix \ref{sec:grav-dual}, we briefly review the gravitational dual of $\cN=2^*$ SYM.} it is convenient to write the mass deformation in terms of relevant operators in irreducible representations of the $\cN = 4$ R-symmetry group, $SO(6) \simeq SU(4)$. There are two contributions that correspond to scalars in the gravitational dual. 

First, there is a dimension-2 {\it bosonic} operator $\cO_2$: 
\beq
\cO_2 =  \frac{1}{3} \Big( \phi_1^{\phantom{\dagger}} \phi_1^\dagger + \phi_2^{\phantom{\dagger}} \phi_2^\dagger - 2 \phi_3^{\phantom{\dagger}} \phi_3^\dagger \Big)~.
\eeq

This operator is an element of the ${\mathbf 20}'$ of $SO(6)$. It contributes the usual positive bosonic mass terms for the hypermultiplet scalars $\phi_1$ and $\phi_2$. But it also destabilizes the scalar $\phi_3$ belonging to the vector multiplet.

The second contribution $\cO_3$ is a dimension-3 {\it fermionic} operator. It introduces mass terms for the Weyl fermions in the hypermultiplet, in addition to trilinear scalar terms and scalar mass terms. The operator $\cO_3$ is
\bea
\cO_3 &=&  2\Big( - \lambda_1^{\phantom{1}\alpha} \lambda_{2\alpha} - \lambdab^1_{\phantom{1}\dot{\alpha}} \lambdab^{2\dot{\alpha}} \nn \\
&& - \rtwo \phi_3^{\phantom{\dagger}}[\phi_1^{\phantom{\dagger}}, \phi_1^\dagger] - \rtwo \phi_3 [\phi_2^{\phantom{\dagger}}, \phi_2^\dagger] - \rtwo \phi_3^\dagger[\phi_1^{\phantom{\dagger}}, \phi_1^\dagger] - \rtwo \phi_3^\dagger [\phi_2^{\phantom{\dagger}}, \phi_2^\dagger]\Big) \nn \\
&& + \frac{2}{3} m \Big(\phi_1^{\phantom{\dagger}} \phi_1^\dagger + \phi_2^{\phantom{\dagger}} \phi_2^\dagger + \phi_3^{\phantom{\dagger}} \phi_3^\dagger \Big)~.
\eea

We also note that $\cO_3$ contains the Konishi operator, which is an $SO(6)$ singlet 
\beq
\cO_K = \sum_{i=1}^3 \phi_i^{\phantom{\dagger}} \phi_i^\dagger~.
\eeq 

This term is crucial since it cancels the negative potential energy for $\phi_3$ introduced by operator $\cO_2$. 

Thus the action of the $\cN = 2^*$ SYM can be expressed as \cite{Hoyos:2011uh, Buchel:2012gw}
\bea
S_{\cN=2^*} &=& S_{\cN=4} - \frac{1}{2g^2} \int d^4x~ m^2~\Tr \cO_2 - \frac{1}{2 g^2} \int d^4x~ m~\Tr \cO_3~.
\eea

In general one could consider the case where the mass parameter is unequal for the two operators. In Ref. \cite{Buchel:2007vy} the authors have explored the thermodynamics of the $\cN = 2^*$ Yang-Mills plasma for a wide range of temperatures and for different mass deformations $(m_{\rm bosonic}, m_{\rm fermionic})$. Supersymmetry is softly broken by the temperature and unequal values of mass parameters in such cases.  

We note that the $\cN = 2^*$ SYM theory has an $SU(2) \times U(1)$ R-symmetry. The symmetry breaking gives equal masses to two of the four Weyl fermions. The $SU(2)$ acts on the two massless fermions, and the $U(1) \simeq SO(2)$ mixes the two massive fermions. As $m \to 0$, we recover $\cN = 4$ SYM theory. When $m \to \infty$, the massive fields decouple from the theory and we end up with four-dimensional Yang-Mills with $\cN = 2$ supersymmetry.

%%%%%%%%%%%%%%%%%%%%%%%%%%%%%%%%%%%%%%%%%%%%%%%%%%%%
\section{Vafa-Witten Twist of $\cN = 4$ SYM}
\label{sec:Vafa-Witten-twist}
%%%%%%%%%%%%%%%%%%%%%%%%%%%%%%%%%%%%%%%%%%%%%%%%%%%%

In this section, we briefly review the Vafa-Witten twist of $\cN = 4$ SYM, which is crucial for the supersymmetric lattice formulation of $\cN = 2^*$ SYM. Since we are interested in formulating $\cN = 2^*$ SYM on a Euclidean spacetime lattice, we begin with $\cN = 4$ SYM on flat ${\mathbb R}^4$.

Four-dimensional $\cN = 4$ SYM can be twisted in three inequivalent ways, giving rise to $(i)$ half-twisted theory \cite{Yamron:1988qc}, $(ii)$ Vafa-Witten theory (gauged four-dimensional A model) \cite{Vafa:1994tf} and $(iii)$ geometric Langlands twisted theory (gauged four-dimensional B model or Marcus twisted theory) \cite{Yamron:1988qc, Marcus:1995mq}. When the theory is formulated on a flat manifold or in general on a hyper-Kahler manifold, the twisted theories coincide with the untwisted $\cN = 4$ SYM theory \cite{Vafa:1994tf}.

For the Vafa-Witten twist of $\cN = 4$ SYM, the internal symmetry group $SU(4)$ is decomposed as $SU(2)_F \times SU(2)_I$ such that the twisted global symmetry group is 
\beq
SU(2)'_L \times SU(2)_R \times SU(2)_F~,
\eeq
where
\beq
SU(2)'_L = {\rm diag}~\Big( SU(2)_L \times SU(2)_I \Big)~,
\eeq
and $SU(2)_F$ remains as a residual internal symmetry group. The fields and supercharges of the untwisted theory are rewritten in terms of the twisted fields. 

After performing the twist, the fields of $\cN = 4$ SYM decompose in the following way \cite{Labastida:1997vq},
\beq
\begin{aligned}
A_\mu &\longrightarrow A_\mu~, \\
\lambda_{u\alpha} &\longrightarrow \eta^i, \chi^i_{\mu\nu}~, \\
\lambdab^u_{\phantom{u}\dot{\alpha}} &\longrightarrow \psi^i_\mu~, \\
\phi_{uv} &\longrightarrow B_{\mu\nu}, \varphi^{ij}~,
\end{aligned}
\eeq 
with $i, j = 1, 2$ representing the indices of the residual internal rotation group $SU(2)_F$ and $\varphi^{ij}$ a symmetric tensor. The fields $\chi^i_{\mu\nu}$ and $B_{\mu\nu}$ are self-dual with respect to the Euclidean Lorentz indices.

We can further split the fields with $SU(2)_F$ indices in the following way:
\beq
\begin{aligned}
\psi^i_\mu &\longrightarrow \psi_\mu, \chi_\mu~, \\
\eta^i &\longrightarrow \eta, \zeta~, \\
\chi^i_{\mu\nu} &\longrightarrow \chi_{\mu\nu}, \psi_{\mu\nu}~, \\
\varphi^{ij} &\longrightarrow (\phi, \phib, C)~.
\end{aligned}
\eeq 

The gauge field $A_\mu$ is a singlet, the fermions $(\eta, \zeta)$, $(\psi_\mu, \chi_\mu)$ and $(\chi_{\mu\nu}, \psi_{\mu\nu})$ form doublets and the scalars $(\phi, \phib, C)$ form a triplet under $SU(2)_F$. 

The theory exhibits flat directions along which the fields $\phi, \phib, C$ commute with each other. Such configurations are given by diagonal matrices up to gauge transformations, in general.

The subset of the twisted fields $(A_\mu, \phi, \phib, \eta, \psi_\mu, \chi_{\mu\nu})$ can be readily recognized as the twisted vector multiplet of the four-dimensional $\cN = 2$ SYM theory (Donaldson-Witten theory) \cite{Witten:1988ze}. The twisted theory contains an $\cN = 2$ hypermultiplet with the field content $(C, B_{\mu\nu}, \zeta, \chi_\mu, \psi_{\mu\nu})$. We make this hypermultiplet massive when we construct the twisted $\cN = 2^*$ SYM theory. A mass deformed version of Vafa-Witten twisted theory was constructed in Ref. \cite{Kato:2011yh} but this does not correspond to the $\cN = 2^*$ SYM theory.

The twisting procedure gives rise to the following twisted supercharges: two scalars ($Q, \tQ$), two vectors ($Q_\mu, \tQ_\mu$) and two self-dual tensors ($Q_{\mu\nu}, \tQ_{\mu\nu}$). All twisted supercharges leave the twisted $\cN = 4$ SYM action invariant. 

We are interested in the scalar supercharges $Q$ and $\tQ$. The twisted theory is invariant under the Cartan subgroup of $SU(2)_F$. We can define a conserved charge in the theory. We call it the $U(1)_R$ charge. In the topological field theory language it is known as the ghost number. 

The scalar supercharges $Q$ and $\tQ$ have opposite $U(1)_R$ charges. In Table \ref{tab:twisted-fields}, we provide the $U(1)_R$ charges, canonical dimensions and nature (even or odd) of the twisted fields of the $\cN = 4$ Yang Mills.

%%%%%%%%%%%%%%%%%%%%%%%%%%%%%%%%%%%%%%%%%%%%%%%%%%%%%%%%
\begin{table}[t]
\renewcommand{\arraystretch}{1.3}%
\centerline{
\begin{tabular}{ | c | c | c | c |}
\hline
field & $U(1)_R$ charge & dimension & nature \\
\hline
\hline
$A_\mu$ & 0 & 1 & {\rm even} \\
\hline
$\phi$ & 2 & 1 & {\rm even} \\
\hline
$\phib$ & $-2$ & 1 & {\rm even} \\
\hline
$C$ & 0 & 1 & {\rm even} \\
\hline
$B_{\mu\nu}$ & 0 & 1 & {\rm even} \\
\hline
$H_\mu$ & 0 & 2 & {\rm even} \\
\hline
$H_{\mu\nu}$ & 0 & 2 & {\rm even} \\
\hline
$\eta$ & $-1$ & $3/2$ & {\rm odd} \\
\hline
$\zeta$ & 1 & $3/2$ & {\rm odd} \\
\hline
$\psi_\mu$ & 1 & $3/2$ & {\rm odd} \\
\hline
$\chi_\mu$ & $-1$ & $3/2$ & {\rm odd} \\
\hline
$\chi_{\mu\nu}$ & $-1$ & $3/2$ & {\rm odd} \\
\hline
$\psi_{\mu\nu}$ & 1 & $3/2$ & {\rm odd} \\
\hline
\end{tabular}
}
\caption{The $U(1)_R$ charges, canonical dimensions and nature (even or odd) of the twisted fields of the $\cN = 4$ SYM theory.
\label{tab:twisted-fields}}
\end{table}
%%%%%%%%%%%%%%%%%%%%%%%%%%%%%%%%%%%%%%%%%%%%%%%%%%%%%%%%

The action of scalar supercharge $Q$ on twisted fields has the following form
\bea
\label{eq:Q-transforms}
 Q A_\mu &=& - \psi_\mu~, \nn \\
 Q C &=& \rtwo \zeta~, \nn \\
 Q \psi_\mu &=& - 2 \rtwo D_\mu \phi~, \nn \\
 Q \zeta &=& - 2 [\phi, C]~, \nn \\
 Q \phi &=& 0~, \nn \\
 Q \phib &=& \rtwo \eta~, \nn \\
 Q \chi_\mu &=& 2 H_\mu~, \\
 Q \eta &=& - 2 [\phi, \phib]~, \nn \\
 Q H_\mu &=& - \rtwo [\phi, \chi_\mu]~, \nn \\
 Q \chi_{\mu\nu} &=& 2 H_{\mu\nu}~, \nn \\
 Q B_{\mu\nu} &=& \rtwo \psi_{\mu\nu}~, \nn \\
 Q H_{\mu\nu} &=& - \rtwo [\phi, \chi_{\mu\nu}]~, \nn \\
 Q \psi_{\mu\nu} &=& - 2 [\phi, B_{\mu\nu}]~, \nn
\eea
where we have introduced two auxiliary fields, a vector field $H_\mu$ and a self-dual tensor $H_{\mu\nu}$.  We have also defined the field strength $F_{\mu\nu}$ and covariant derivative $D_\mu$ the following way
\beq
F_{\mu\nu} \equiv [D_\mu, D_\nu]~,~~D_\mu \equiv \partial_\mu + [A_\mu, ~~]~.
\eeq

We note that the $Q$ supercharge satisfies the following algebra,
\beq
\begin{aligned}
	Q^2 A_\mu &= 2 \rtwo D_\mu \phi~, \\
	Q^2 X &= 2 \rtwo [X, \phi]~,
\end{aligned}
\eeq
for a generic field $X$. $Q$ is nilpotent up to infinitesimal gauge transformations with parameter $\phi$.

The $\tQ$ supercharge acts on the twisted fields in the following way: 
\bea
\label{eq:q-tilde-transforms}
 \tQ A_\mu &=& - \chi_\mu~, \nn \\
 \tQ C &=& - \rtwo \eta~, \nn \\
 \tQ \chi_\mu &=& 2 \rtwo D_\mu \phib~, \nn \\
 \tQ \eta &=& - 2 [\phib, C]~, \nn \\
 \tQ \phib &=& 0~, \nn \\
 \tQ \psi_\mu &=& - 2 H_\mu + 2 \rtwo D_\mu C~, \nn \\
 \tQ \phi &=& \rtwo \zeta~, \\
 \tQ H_\mu &=& - \rtwo [\phib, \psi_\mu] - \rtwo [\chi_\mu, C] - 2 D_\mu \eta~, \nn \\
 \tQ \zeta &=& 2 [\phib, \phi]~,  \nn \\
 \tQ B_{\mu\nu} &=& - \rtwo \chi_{\mu\nu}~, \nn \\
 \tQ \psi_{\mu\nu} &=& 2 H_{\mu\nu} + 2 [C, B_{\mu\nu}]~, \nn \\
 \tQ \chi_{\mu\nu} &=& - 2 [\phib, B_{\mu\nu}]~, \nn \\
 \tQ H_{\mu\nu} &=& \rtwo [\phib, \psi_{\mu\nu}] + \rtwo [\eta, B_{\mu\nu}] + \rtwo [C, \chi_{\mu\nu}]~. \nn 
\eea

The $\tQ$ supercharge satisfies the following algebra,
\beq
\label{eq:tQ-closure}
\begin{aligned}
	\tQ^2 A_\mu &= - 2 \rtwo D_\mu \phib~, \\
	\tQ^2 X &= - 2 \rtwo [X, \phib]~,
\end{aligned}
\eeq
for a generic field $X$. $\tQ$ is nilpotent up to infinitesimal gauge transformations with parameter $\phib$.

We can obtain the twisted action of the $\cN = 4$ SYM theory through successive variations of $Q$ and $\tQ$ on a functional $\cF$ known as the action potential \cite{Labastida:1997vq, Labastida:1997xk}. We have the twisted action
\beq
S_{\cN=4} = Q \tQ ~\frac{1}{g^2} \int d^4x~\cF~,
\eeq
where
\bea
\label{eq:action-pot}
\cF &=& \Tr \Big( -\frac{1}{2 \rtwo} B_{\mu\nu} F_{\mu\nu} - \frac{1}{24 \rtwo} B_{\mu\nu} [B_{\mu\rho}, B_{\nu\rho}] \nn \\
&&~~~~~~~~~~~~~~~~~~~~~~~~~~~~~~~~~~~~~~ - \frac{1}{8} \chi_{\mu\nu} \psi_{\mu\nu} - \frac{1}{8} \psi_\mu \chi_\mu - \frac{1}{8} \eta \zeta \Big)~.   
\eea

The Vafa-Witten twisted action can be written as the $Q$ variation of a gauge fermion $\Psi$ (which in turn is the $\tQ$ variation of $\cF$)
\bea
S_{\cN=4} &=& Q \frac{1}{g^2} \int d^4x~ \Psi~,
\eea
with $\Psi$ taking the form
\bea
\label{eq:gf}
\Psi &=& \Tr \Big( \chi_{\mu\nu} \Big[\hf F_{\mu\nu} + \qtr H_{\mu\nu} + \frac{1}{8} [B_{\mu\rho}, B_{\nu\rho}] + \qtr [C, B_{\mu\nu}] \Big] \nn \\
&& \quad \quad + \frac{1}{2\rtwo} \psi_\mu (D_\mu \phib) - \qtr \eta [\phi, \phib] - \qtr \zeta [C, \phib] - \frac{1}{4} \psi_{\mu\nu} [B_{\mu\nu}, \phib] \nn \\
&& \quad \quad + \chi_\mu \Big[ \qtr H_\mu - \frac{1}{2\rtwo} (D_\mu C) - \frac{1}{2 \rtwo} (D_\nu B_{\nu\mu}) \Big] \Big)~.   
\eea

Applying $Q$ variation on the gauge fermion, Eq. (\ref{eq:gf}), we obtain the twisted $\cN = 4$ SYM action
\bea
\label{eq:twisted-action}
S_{\cN=4} &=& \frac{1}{g^2} \int d^4x~\Tr \Big( H_{\mu\nu} \Big[ F_{\mu\nu} + \hf H_{\mu\nu} + \qtr [B_{\mu\rho}, B_{\nu\rho}] + \hf [C, B_{\mu\nu}] \Big] \nn \\
&& \quad \quad \quad - (D_\mu \phi) (D_\mu \phib) + \hf [\phi, \phib]^2 - \hf [\phib, C] [\phi, C] - \hf [\phi, B_{\mu\nu}] [\phib, B_{\mu\nu}] \nn \\
&& \quad \quad \quad + H_\mu \Big[ \hf H_\mu - \frac{1}{\rtwo} (D_\mu C) - \frac{1}{\rtwo} (D_\nu B_{\nu\mu}) \Big] \nn \\
&& \quad \quad \quad + \hf \chi_{\mu\nu} (D_\mu \psi_\nu) + \hf \psi_{\mu\nu} (D_\mu \chi_\nu) - \hf \psi_\mu (D_\mu \eta) + \hf \chi_\mu (D_\mu \zeta) \nn \\
&& \quad \quad \quad + \frac{1}{2\rtwo} \eta [\phi, \eta] + \frac{1}{2\rtwo} \chi_\mu [\phi, \chi_\mu] + \frac{1}{2\rtwo} \chi_{\mu\nu} [\phi, \chi_{\mu\nu}] \nn \\
&& \quad \quad \quad - \frac{1}{2\rtwo} \zeta [\phib, \zeta] - \frac{1}{2\rtwo} \psi_\mu [\phib, \psi_\mu] - \frac{1}{2\rtwo} \psi_{\mu\nu} [\phib, \psi_{\mu\nu}] \nn \\
&& \quad \quad \quad - \frac{1}{2\rtwo} \eta [\zeta, C] - \frac{1}{2\rtwo} \chi_\mu [\psi_\mu, C] + \frac{1}{2\rtwo} \chi_{\mu\nu} [\psi_{\mu\nu}, C] \nn \\
&& \quad \quad \quad - \frac{1}{2 \rtwo} \psi_\mu [\chi_\nu, B_{\mu\nu}] - \frac{1}{2\rtwo} \chi_{\mu\nu} [\psi_{\mu\rho}, B_{\nu\rho}] \nn \\
&& \quad \quad \quad - \frac{1}{2\rtwo} \chi_{\mu\nu} [\zeta, B_{\mu\nu}] - \frac{1}{2\rtwo} \psi_{\mu\nu} [\eta, B_{\mu\nu}] \Big)~.
\eea

The twisted action given above has net $U(1)_R$ charge zero. 

%%%%%%%%%%%%%%%%%%%%%%%%%%%%%%%%%%%%%%%%%%%%%%%%%%%%%%%%%%%
\section{$\cN = 2^*$ SYM using Twisted Fields}
\label{sec:twisted-n2*-sym}
%%%%%%%%%%%%%%%%%%%%%%%%%%%%%%%%%%%%%%%%%%%%%%%%%%%%%%%%%%%

After reviewing the existing literature on Vafa-Witten twisted $\cN=4$ SYM we are at a point to write down the twisted action of $\cN=2^*$ SYM. Once we know the transformations from the untwisted fields to twisted fields it is straightforward to write down the action of the $\cN = 2^*$ SYM theory in the twisted language. The $\cN = 2^*$ SYM theory is obtained by giving masses to the $\cN = 2$ hypermultiplet fields $(C, B_{\mu\nu}, \zeta, \chi_\mu, \psi_{\mu\nu})$.

We can rewrite the mass terms given in Eq. (\ref{eq:n2-star-untwisted-mass}) using the twisted fields. In the bosonic sector, the components of the untwisted fields are related to that of the twisted fields the following way \cite{Labastida:1997vq, Labastida:1997xk} [note that we use the anti-Hermitian basis for $SU(N)$ generators]:
\begin{align}
\label{eq:phi-i-to-twisted}
 \phi_1^{\phantom{\dagger}}  &= \frac{1}{\sqrt{2}} (B_{13} + i C)~,   			&   \phi_1^\dagger  &= \frac{1}{\sqrt{2}} (-B_{13} + i C)~, \nn \\
 \phi_2^{\phantom{\dagger}}  &= \frac{1}{\sqrt{2}} (B_{12} + i B_{23})~,   	&   \phi_2^\dagger  &= \frac{1}{\sqrt{2}} (-B_{12} + i B_{23})~, \\
 \phi_3^{\phantom{\dagger}}  &= - \frac{1}{\sqrt{2}} \phib~,   								&   \phi_3^\dagger  &= - \frac{1}{\sqrt{2}} \phi~. \nn
\end{align}

Substituting the twisted field variables, we obtain the bosonic mass terms and the trilinear coupling terms
\bea
\label{eq:boson-mass}
m^2 \phi_1^{\phantom{\dagger}} \phi_1^\dagger + m^2 \phi_2^{\phantom{\dagger}} \phi_2^\dagger &=& - \hf m^2 B_{\mu\nu}^2 - \hf m^2 C^2~, \nn \\
- \rtwo m \phi_3^{\phantom{\dagger}}[\phi_1^{\phantom{\dagger}}, \phi_1^\dagger] - \rtwo m \phi_3 [\phi_2^{\phantom{\dagger}}, \phi_2^\dagger] &=& - \frac{1}{2} m \phib [B_{\mu\nu}, B_{\mu\nu}] - \frac{1}{2} m \phib [C, C] \nn \\
&& \quad \quad + i m \phib \Big([B_{12}, B_{23}] + [B_{13}, C] \Big)~,  \\
- \rtwo m \phi_3^\dagger[\phi_1^{\phantom{\dagger}}, \phi_1^\dagger] - \rtwo m \phi_3^\dagger [\phi_2^{\phantom{\dagger}}, \phi_2^\dagger] &=& - \frac{1}{2} m \phi [B_{\mu\nu}, B_{\mu\nu}] - \frac{1}{2} m \phi [C, C] \nn \\
&& \quad \quad + i m \phi \Big([B_{12}, B_{23}] + [B_{13}, C]\Big)~. \nn
\eea

In the fermionic sector, we have the following relations between the twisted and untwisted field variables \cite{Labastida:1997vq, Labastida:1997xk} (we use the conventions given in Ref. \cite{McKeon:2001pm} for Euclidean spinors):
\begin{align}
 \lambda_1^{\phantom{1}1} &= \psi_{12} + \frac{i}{2\rtwo} \psi_{23}~, & \lambda_{21} &= \psi_{12} - \frac{i}{2\rtwo} \psi_{23}~, \nn \\
 \lambda_1^{\phantom{1}2} &= \psi_{13} + \frac{i}{2\rtwo} \zeta~, & \lambda_{22} &= \psi_{13} - \frac{i}{2\rtwo} \zeta~, \\
 \lambdab_1^{\phantom{1}\dot{\alpha}} &= \chi_{2\dot{\alpha}}~, & \lambdab^2_{\phantom{1}\dot{\alpha}} &= - \frac{1}{2\rtwo}\chi_{1\dot{\alpha}}~. \nn
\end{align}

From the above relations we can write down the fermionic mass terms in the language of twisted fields. (In Appendix \ref{sec:spinor-mass}, we derive the fermion mass terms in twisted form.)

The mass terms take the form
\bea
- m ~\Tr \lambda_1^{\phantom{1}\alpha} \lambda_{2\alpha} &=& \frac{im}{\rtwo} ~\Tr (\psi_{12}\psi_{23} + \psi_{13}\zeta)~, \\ 
- m ~\Tr \Big(\lambdab^1_{\phantom{1}\dot{\alpha}} \lambdab^{2\dot{\alpha}}\Big) &=& - \frac{im}{\rtwo} ~\Tr (\chi_1 \chi_2 - \chi_3 \chi_4)~.
\eea

Having expressed the mass deformation terms using twisted variables, it is now straightforward to write down the twisted action of the $\cN = 2^*$ SYM. We have
\bea
\label{eq:n2-star-action}
S_{\cN=2^*} &=& S_{\cN=4} + S_m~,
\eea
where $S_{\cN=4}$ is given in Eq. (\ref{eq:twisted-action}) and $S_m$ has the form
\bea
S_m &=& \frac{1}{g^2} \int d^4x~ \Tr \Big[ - \hf m^2 B_{\mu\nu}^2 - \hf m^2 C^2 \nn \\
&& \quad \quad - \hf m \phi \Big([B_{\mu\nu}, B_{\mu\nu}] + [C, C] \Big) - \hf m \phib \Big([B_{\mu\nu}, B_{\mu\nu}] + [C, C] \Big) \nn \\
&& \quad \quad + i m \phi \Big([B_{12}, B_{23}] + [B_{13}, C]\Big) + i m \phib \Big([B_{12}, B_{23}] + [B_{13}, C]\Big) \nn \\
&& \quad \quad + \frac{im}{\rtwo} (\psi_{12}\psi_{23} + \psi_{13} \zeta) - \frac{im}{\rtwo} (\chi_1 \chi_2 - \chi_3 \chi_4) \Big]~.
\eea

We note that the above form of the twisted $\cN = 2^*$ SYM action exhibits the following interesting properties:

\begin{enumerate}
\item[1.] There are terms in $S_m$ with nonzero $U(1)_R$ charges.

We note that $S_m$ contains terms with $U(1)_R$ charge $-2$, $0$ and $+2$, while the $\cN=4$ SYM part of the action contains only terms with $U(1)_R$ charge 0. 

\item[2.] There exist mass terms that are not invariant under twisted Lorentz symmetry. 

We note that there are terms in $S_m$ that contain uncontracted twisted Lorentz indices. The presence of such terms is expected from the fact that the mass deformation of $\cN=4$ SYM breaks R-symmetry from $SU(4)$ to $SU(2) \times U(1)$. However,  we note that the presence of twisted Lorentz symmetry breaking terms does not lead to any inconsistencies in the formulation. The theory is still Lorentz invariant. A consequence of twisted Lorentz symmetry breaking would be the appearance of additional counterterms in the lattice version of the theory.
\end{enumerate}

It would be interesting to ask if we could write down the twisted action of $\cN = 2^*$ SYM in a $Q$-exact form, with an appropriate gauge fermion. In order to achieve this, we need to modify the $Q$ and $\tQ$ transformations on the twisted fields. Let us define $Q^{(m)}$, $\tQ^{(m)}$ and $\Psi^{(m)}$ as the modified scalar supercharges and gauge fermion, respectively.

The action of $\tQ^{(m)}$ on twisted fields is the same as Eq. (\ref{eq:q-tilde-transforms}) except for the fields $\eta$, $H_\mu$ and $\chi_{\mu\nu}$. We have
\bea
\tQ^{(m)} \eta &=& - 2 [\phib, C] + 2 m C~, \nn \\
\tQ^{(m)} H_\mu &=& - \rtwo [\phib, \psi_\mu] - \rtwo [\chi_\mu, C] - 2 D_\mu \eta + \rtwo m \psi_\mu~, \\
\tQ^{(m)} \chi_{\mu\nu} &=& - 2 [\phib, B_{\mu\nu}] - 2 m B_{\mu\nu}~. \nn
\eea

We can show that the $\tQ^{(m)}$ transformations respect the following modified algebra
\beq
\label{eq:tQ-closure-mass}
\begin{aligned}
	(\tQ^{(m)})^2 A_\mu &= - 2 \rtwo D_\mu \phib~, \\
	(\tQ^{(m)})^2 X &= - 2 \rtwo [X, \phib] - 2\rtwo m \alpha X~,
\end{aligned}
\eeq
for a generic field $X$; with $\alpha = 1$ for the fields $(\eta, \psi_\mu, C, H_\mu)$, $\alpha = -1$ for the fields $(\chi_{\mu\nu}, B_{\mu\nu})$ and $\alpha = 0$ for the rest of the fields.

The $Q^{(m)}$ transformations on the twisted fields are the same as the ones given in Eq. (\ref{eq:Q-transforms}) except for the fields $\zeta$, $H_\mu$ and $\psi_{\mu\nu}$. The transformations on these fields are modified in the following way: 
\bea
Q^{(m)} \zeta &=& - 2 [\phi, C] + 2 m C~, \nn \\
Q^{(m)} H_\mu &=& - \rtwo [\phi, \chi_\mu] + \rtwo m \chi_\mu~,  \\
Q^{(m)} \psi_{\mu\nu} &=& - 2 [\phi, B_{\mu\nu}] + 2 m B_{\mu\nu}~. \nn
\eea

We can show that the $Q^{(m)}$ supercharge satisfies the following algebra.
\beq
\label{eq:Q-closure-mass}
\begin{aligned}
	(Q^{(m)})^2 A_\mu &= 2 \rtwo D_\mu \phi~, \\
	((Q^{(m)})^2 X &= 2 \rtwo [X, \phi] + 2 \rtwo m \alpha X~,
\end{aligned}
\eeq
for a generic field $X$, with $\alpha = 1$ for the fields $\zeta, \chi_\mu, \psi_{\mu\nu}, C, H_\mu, B_{\mu\nu}$ and $\alpha = 0$ for the rest of the fields.

It would be interesting to see if the deformation part of the algebra represents rotation by an R-symmetry generator. Similar topics were considered in Ref. \cite{Hanada:2010kt} by Hanada {\it et al.}, and they were extended to various cases by Kato {\it et al.} in Ref. \cite{Kato:2011yh}. It would be interesting to find the structure of the mass deformed supersymmetry algebra shown above.  

In order to derive the twisted Lorentz noninvariant part of the $\cN = 2^*$ SYM action, let us consider linear combinations of the massive fields. The $Q^{(m)}$ transformations give
\beq
\label{eq:Q-on-L-ninv}
\begin{aligned}
Q^{(m)} (\chi_1 + i \chi_2) &= 2 (H_1 + i H_2)~, \\
Q^{(m)} (\chi_3 + i \chi_4) &= 2 (H_3 + i H_4)~, \\
Q^{(m)} (H_1 + i H_2) &= - \rtwo [\phi, (\chi_1 + i \chi_2)] + \rtwo m (\chi_1 + i \chi_2)~, \\
Q^{(m)} (H_3 + i H_4) &= - \rtwo [\phi, (\chi_3 + i \chi_4)] + \rtwo m (\chi_3 + i \chi_4)~, \\
Q^{(m)} (B_{12} + i B_{23}) &= \rtwo (\psi_{12} + i \psi_{23})~, \\
Q^{(m)} (B_{13} + i C) &= \rtwo (\psi_{13} + i \zeta)~, \\
Q^{(m)} (\psi_{12} + i \psi_{23}) &= - 2 [\phi, (B_{12} + i B_{23})] + 2 m (B_{12} + i B_{23})~, \\
Q^{(m)} (\psi_{13} + i \zeta) &= - 2 [\phi, (B_{13} + i C)] + 2 m (B_{13} + iC)~. 
\end{aligned}
\eeq

We can now obtain the $\cN = 2^*$ SYM action as a $Q^{(m)}$ variation of the following modified gauge fermion,
\bea
\Psi^{(m)} &=& \Tr \Big(\chi_{\mu\nu} \Big[\hf F_{\mu\nu} - \qtr H_{\mu\nu} - \frac{1}{8} [B_{\mu\rho}, B_{\nu\rho}] - \frac{1}{4} [C, B_{\mu\nu}] \Big] \nn \\
&& \quad \quad + \frac{1}{2\rtwo} \psi_\mu (D_\mu \phib) - \frac{1}{4} \eta [\phi, \phib] + (\cV + \cW + \cY) \nn \\
&& \quad \quad - \qtr \zeta [C, \phib] - \qtr \psi_{\mu\nu} [B_{\mu\nu}, \phib] + \cT \nn \\
&&\quad \quad + \chi_\mu \Big[- \frac{1}{2 \rtwo} (D_\mu C) - \frac{1}{2\rtwo} (D_\nu B_{\nu\mu}) \Big] \Big)~,
\eea
where
\bea
\cV &=& -\qtr m \Big( (\psi_{12} - i \psi_{23}) (B_{12} + i B_{23}) + (\psi_{13} - i \zeta)(B_{13} + i C)\Big)~, \\
\cW &=& \frac{i}{4} \Big( - \psi_{12} [\phib, B_{23}] + \psi_{23} [\phib, B_{12}] + \eta [B_{12}, B_{23}] \Big)~, \\
\cY &=& \frac{i}{4} \Big( - \psi_{13} [\phib, C] + \zeta [\phib, B_{13}] + \eta [B_{13}, C] \Big)~, \\
\cT &=& \qtr \Big( (\chi_1 - i \chi_2)(H_1 + iH_2) + (\chi_3 + i \chi_4)(H_3 - iH_4) \Big)~.
\eea

We derive the $Q^{(m)}$ transformations of the gauge fermion components $\cV, \cW, \cY$ and $\cT$ in Appendix \ref{sec:ELN-terms}. We also note that the terms $\cW, \cY$ and $\cT$ contain the rotated fields that give appropriate twisted Lorentz noninvariant mass terms of the theory.

It is straightforward to show that $Q^{(m)}$ variation of $\Psi^{(m)}$ will produce the twisted action of $\cN = 2^*$ SYM given in Eq. (\ref{eq:n2-star-action}). Thus, we have
\beq
S_{\cN=2^*} = \frac{1}{g^2} \int d^4x~ Q^{(m)}\Psi^{(m)}~.
\eeq

We note that we are unable to express the twisted action of $\cN=2^*$ SYM as successive variations of $Q^{(m)}$ and $\tQ^{(m)}$ on an action functional, say, $\cF^{(m)}$. This could be due to the fact that we are interested in giving mass to the $\cN=2$ multiplet associated with the supercharge $Q^{(m)}$, that is, for the multiplet with the twisted fields $(C, B_{\mu\nu}, \zeta, \chi_\mu, \psi_{\mu\nu})$, where the fermions $\zeta$, $\chi_\mu$ and $\psi_{\mu\nu}$ are originally associated with the twisted supercharges $Q$, $Q_\mu$ and $Q_{\mu\nu}$, respectively.

%%%%%%%%%%%%%%%%%%%%%%%%%%%%%%%%%%%%%%%%%%%%%%%%%%%
\section{Lattice Formulation}
\label{sec:lattice}
%%%%%%%%%%%%%%%%%%%%%%%%%%%%%%%%%%%%%%%%%%%%%%%%%%%

%%%%%%%%%%%%%%%%%%%%%%%%%%%%%%%%%%%%%%%%%%%%%%%%%%%
\subsection{Balanced Topological Field Theory Form}
%%%%%%%%%%%%%%%%%%%%%%%%%%%%%%%%%%%%%%%%%%%%%%%%%%%

We can rewrite the Vafa-Witten twisted $\cN = 4$ SYM theory in a form known as the balanced topological field theory form. The existence of two scalar supercharges $Q$ and $\tQ$ would allow us to express the $\cN = 4$ SYM theory in this form. In Ref. \cite{Dijkgraaf:1996tz} Dijkgraf and Moore wrote down the BTFT form of the Vafa-Witten twisted theory. Sugino used this approach to formulate four-dimensional $\cN = 4$ and $\cN = 2$ SYM theories on the lattice \cite{Sugino:2003yb}.

We can define a three-component vector $\vec{\Phi}$, which is a function of the field strength. The components of this vector take the form
\beq
\Phi_{\tt A} \equiv 2 \Big( F_{\tt A4} + \hf \epsilon_{\tt ABC} F_{\tt BC}\Big)~,
\eeq
with ${\tt A}, {\tt B}, {\tt C} = \boldsymbol{1}, \boldsymbol{2}, \boldsymbol{3}$. Similarly, we introduce three-component vector fields $\vec{B}$, $\vec{H}$, $\vec{\psi}$ and $\vec{\chi}$. 

The action potential takes the following form in the BTFT notation
\bea
\cF &=& \Big( - \frac{1}{2 \rtwo} B_{\tt A} \Phi_{\tt A} - \frac{1}{24 \rtwo} \epsilon_{\tt ABC} B_{\tt A} [B_{\tt B}, B_{\tt C}] - \frac{1}{8} \chi_{\tt A} \psi_{\tt A} - \frac{1}{8} \psi_\mu \chi_\mu - \frac{1}{8} \eta \zeta \Big)~.
\eea

It is straightforward to write down the $\tQ^{(m)}$ and $Q^{(m)}$ transformations on the twisted fields in BTFT form. In particular, the $Q^{(m)}$ transformations take the form 
\bea
Q^{(m)} A_\mu &=& - \psi_\mu~, \nn \\
Q^{(m)} \psi_\mu &=& - 2 \rtwo D_\mu \phi~, \nn \\
Q^{(m)} \phi  &=& 0~, \nn \\
Q^{(m)} \phib &=& \rtwo \eta~, \nn \\
Q^{(m)} \eta &=& - 2 [\phi, \phib]~, \nn \\
Q^{(m)} C &=& \rtwo \zeta~, \nn \\
Q^{(m)} \zeta &=& - 2 [\phi, C] + 2 m C~, \\
Q^{(m)} \chi_\mu &=& 2 H_\mu~, \nn \\
Q^{(m)} H_\mu &=& - \rtwo [\phi, \chi_\mu] + \rtwo m \chi_\mu~, \nn \\
Q^{(m)} B_{\tt A} &=& \rtwo \psi_{\tt A}~, \nn \\
Q^{(m)} \psi_{\tt A} &=& - 2 [\phi, B_{\tt A}] + 2 m B_{\tt A}~, \nn \\
Q^{(m)} \chi_{\tt A} &=& 2 H_{\tt A}~, \nn \\
Q^{(m)} H_{\tt A} &=& - \rtwo [\phi, \chi_{\tt A}]~. \nn
\eea

The gauge fermion has the following BTFT form
\bea
\Psi_{\rm BTFT}^{(m)} &=&  \Tr \Big(\chi_{\tt A} \Big[\hf F_{\tt A} - \qtr H_{\tt A} - \frac{1}{8} \epsilon_{\tt ABC}[B_{\tt B}, B_{\tt C}] - \frac{1}{4} [C, B_{\tt A}] \Big] \nn \\
&& \quad \quad + \frac{1}{2\rtwo} \psi_\mu (D_\mu \phib) - \frac{1}{4} \eta [\phi, \phib] + (\cV + \cW + \cY) \nn \\
&& \quad \quad - \qtr \zeta [C, \phib] - \qtr \psi_{\tt A} [B_{\tt A}, \phib] + \cT \nn \\
&&\quad \quad - \frac{1}{2 \rtwo} \chi_\mu (D_\mu C) + \frac{1}{2\rtwo} B_{\tt A} (D \chi)_{\tt A} \Big)~,
\eea
where
\bea
\cV &=& -\qtr m \Big( (\psi_{\boldsymbol{3}} - i \psi_{\boldsymbol{1}}) (B_{\boldsymbol{3}} + i B_{\boldsymbol{1}}) + (\psi_{\boldsymbol{2}} - i \zeta)(B_{\boldsymbol{2}} + i C)\Big)~, \\
\cW &=& \frac{i}{4} \Big( - \psi_{\boldsymbol{3}} [\phib, B_{\boldsymbol{1}}] + \psi_{\boldsymbol{1}} [\phib, B_{\boldsymbol{3}}] + \eta [B_{\boldsymbol{3}}, B_{\boldsymbol{1}}] \Big)~, \\
\cY &=& \frac{i}{4} \Big( - \psi_{\boldsymbol{2}} [\phib, C] + \zeta [\phib, B_{\boldsymbol{2}}] + \eta [B_{\boldsymbol{2}}, C] \Big)~, \\
\cT &=& \qtr \Big( (\chi_1 - i \chi_2)(H_1 + iH_2) + (\chi_3 + i \chi_4)(H_3 - iH_4) \Big)~,
\eea
and 
\beq
(D \chi)_{\tt A} \equiv 2 \Big( D_{\tt A} \chi_{\tt 4} + \hf \epsilon_{\tt ABC} D_{\tt B} \chi_{\tt C} \Big)~.
\eeq

The action of twisted $\cN = 2^*$ SYM can again be written as $Q^{(m)}$ variation of gauge fermion expressed in BTFT form
\beq
S_{\cN=2^*} = \frac{1}{g^2} \int d^4x~ \Tr Q^{(m)} \Psi_{\rm BTFT}^{(m)}~.
\eeq

%%%%%%%%%%%%%%%%%%%%%%%%%%%%%%%%%%%%%%%%%%%%%%%%%%%
\subsection{Lattice Regularized Theory}
%%%%%%%%%%%%%%%%%%%%%%%%%%%%%%%%%%%%%%%%%%%%%%%%%%%

We formulate the theory on a four-dimensional hypercubic lattice by distributing the degrees of freedom of the theory appropriately on the unit cell of the lattice. It is important that the resulting lattice theory is gauge invariant as we map the continuum fields to corresponding lattice fields. We need to choose an appropriate discretization procedure. We closely follow the discretization prescription given by Sugino in Ref. \cite{Sugino:2003yb}. There exists another discretization prescription, known as the geometric discretization. However, it is not appropriate for the theory we have, since the lattice theory would contain terms that are not gauge invariant if we use the geometric discretization scheme. This is also the reason we did not choose the B-model twist for constructing twisted $\cN=2^*$ SYM. It is impossible to construct gauge invariant mass terms of twisted $\cN=2^*$ SYM on the lattice using the geometric discretization prescription.

We begin by promoting the gauge fields $A_\mu$ to compact unitary variables on the lattice
\bea
U_\mu(\bn) &\equiv& U(\bn, \bn + \mu) = e^{A_\mu(\bn)}~, \\
U^\dagger_\mu(\bn - \mu) &\equiv& U(\bn, \bn - \mu) = e^{-A_\mu(\bn)}~,
\eea
living on the oriented links connecting from site $\bn$ to site $\bn+\mu$ and form site $\bn$ to site $\bn-\mu$, respectively. All other field variables are distributed on the sites under this discretization prescription.

Upon using the language of the BTFT form, we have the $Q^{(m)}$ transformations on the lattice, which are almost the same as their continuum cousins
\begin{align}
Q^{(m)} U_\mu(\bn) &= - \psi_\mu U_\mu(\bn)~,   & Q^{(m)} \psi_\mu (\bn) &= \psi_\mu(\bn) \psi_\mu(\bn) - 2 \rtwo D_\mu^{(+)} \phi(\bn)~, \nn \\
Q^{(m)} \phi(\bn)  &= 0~,          & \nn \\
Q^{(m)} \phib(\bn) &= \rtwo \eta(\bn)~, & Q^{(m)} \eta(\bn) &= - 2 [\phi(\bn), \phib(\bn)]~, \nn \\
Q^{(m)} C(\bn) &= \rtwo \zeta(\bn)~, & Q^{(m)} \zeta(\bn) &= - 2 [\phi(\bn), C(\bn)] + 2 m C(\bn)~, \\
Q^{(m)} \chi_\mu(\bn) &= 2 H_\mu(\bn)~, & Q^{(m)} H_\mu(\bn) &= - \rtwo [\phi(\bn), \chi_\mu(\bn)] + \rtwo m \chi_\mu(\bn)~, \nn \\
Q^{(m)} B_{\tt A}(\bn) &= \rtwo \psi_{\tt A}(\bn)~, & Q^{(m)} \psi_{\tt A}(\bn) &= - 2 [\phi(\bn), B_{\tt A}(\bn)] + 2 m B_{\tt A}(\bn)~, \nn \\
Q^{(m)} \chi_{\tt A}(\bn) &= 2 H_{\tt A}(\bn)~, & Q^{(m)} H_{\tt A}(\bn) &= - \rtwo [\phi(\bn), \chi_{\tt A}(\bn)]~. \nn
\end{align}

These transformations were originally proposed by Sugino in Ref. \cite{Sugino:2003yb} while formulating the $\cN = 4$ and $\cN = 2$ SYM theories on the lattice. 

In the above transformations, $D_\mu^{(+)}$ is the forward covariant difference operator
\beq
D_\mu^{(+)} f(\bn) = U_\mu(\bn) f(\bn + \mu) U^\dagger_\mu(\bn) - f(\bn)~, 
\eeq
and $D_\mu^{(-)}$ represents the backward difference operator
\beq
D^{(-)}_\mu g_\mu(\bn) = g_\mu(\bn) - U^\dagger_\mu(\bn - \mu) g_\mu(\bn - \mu) U_\mu(\bn - \mu)~.
\eeq

The $Q^{(m)}$ transformations reduce to their continuum counterparts in the limit of vanishing lattice spacing. The term quadratic in $\psi_\mu$ is suppressed by additional power of the lattice spacing. $\left(Q^{(m)}\right)^2$ on the lattice obeys a relation similar to the one given in the continuum.

Once we have the $Q^{(m)}$ transformation rule closed among lattice variables, it is almost straightforward to construct the lattice action. 

The functional $\Phi_{\tt A}$ takes the following form on the lattice \cite{Sugino:2003yb}:
\beq
\Phi_{\tt A} (\bn) = - \Big(U_{\tt A4} (\bn) - U_{\boldsymbol{4} {\tt A}}(\bn) + \hf \sum_{ {\tt B, C} = {\boldsymbol{1}} }^{\boldsymbol{3}} \epsilon_{\tt ABC} (U_{\tt BC} (\bn) - U_{\tt CB} (\bn))\Big)~.
\eeq

The plaquette variables $U_{\mu\nu}(x)$ are defined as
\beq
U_{\mu\nu}(\bn) \equiv U_\mu(\bn) U_\nu(\bn + \mu) U_\mu(\bn + \nu)^\dagger U_\nu(\bn)^\dagger~.
\eeq

We can integrate out the auxiliary field $\vec{H}(\bn)$ so that the $\vec{\Phi}(\bn)^2$ term gives the gauge kinetic term on the lattice,
\beq
\label{eq:double-winding}
\frac{1}{2 g_0^2} \sum_\bn \sum_{\mu < \nu} \Tr \Big[ -(U_{\mu\nu}(\bn) - U_{\nu\mu}(\bn))^2 \Big]~.
\eeq

We note that there are also additional terms appearing in $\vec{\Phi}(\bn)^2$ as cross terms. They become topological (total derivative) terms in the continuum limit; however, we should keep them at the lattice level. The gauge terms in the continuum are $\left(F_{\mu\nu} + \widetilde{F}_{\mu\nu}\right)^2$ rather than conventional $F_{\mu\nu}^2$. The vacua in the continuum theory are instanton solutions (anti-self-dual field strengths) corresponding to $\Phi_{\tt A} = 0$. 

We note that the above term (\ref{eq:double-winding}) contains double winding plaquette terms. On the other hand, the standard Wilson action has the form
\beq
\frac{1}{2 g_0^2} \sum_\bn \sum_{\mu < \nu} \Tr \Big[ 2 - U_{\mu\nu}(\bn) - U_{\nu\mu}(\bn) \Big]~,
\eeq
which has a unique minimum $U_{\mu\nu} = {\mathbb I}$. 

The action obtained through discretizing the twisted theory this way has many classical vacua 
\beq
\label{eq:many-classical-vacua}
U_{\mu\nu} = {\rm diag} (\pm 1, \cdots, \pm 1)~,
\eeq
up to gauge transformations, where any combinations of $\pm 1$, with $-1$ appearing even times, are allowed in the diagonal entries. 

We also note that in the case of $G = SU(N)$, in addition to Eq. (\ref{eq:many-classical-vacua}), there also appear the center elements
\beq
\label{eq:many-classical-vacua-2}
U_{\mu\nu} = z_k {\mathbb I}_N = \exp(2 \pi i k /N)~{\rm diag}(1, 1, \cdots, 1)~~~~(k = 1, 2, \cdots, N-1)~
\eeq
as the minima.

The existence of many classical vacua has some serious consequences. Since the diagonal entries can be taken freely for each plaquette, it results in a huge degeneracy of vacua with the number growing as exponential of the number of plaquettes. We need to add up contributions from all the minima in order to see the dynamics of the model. In this case, the ordinary weak field expansion around a single vacuum $U_{\mu\nu} = {\mathbb I}$ cannot be justified. That is, we are unable to say anything about the continuum limit of the lattice theory without its nonperturbative investigations.  

We could add a term proportional to the standard Wilson action to the lattice action in order to resolve the degeneracy
\beq
\label{eq:Wilson-term}
\Delta S = \frac{\rho}{2 g_0^2} \sum_\bn \sum_{\mu < \nu} \Tr \Big[2 - U_{\mu\nu}(\bn) - U_{\nu\mu}(\bn)\Big]~,
\eeq
where $\rho$ is a parameter to be tuned. This term resolves the degeneracy with the split $4 \rho/g_0^2$ \cite{Sugino:2003yb}.

We note that this breaks the supersymmetry $Q^{(m)}$, even though it justifies the expansion around the vacuum $U_{\mu\nu} = {\mathbb I}$.

On the lattice, we have a lattice version of the anti-self-dual equations for the minima. A discussion about lattice anti-self-dual equations is lacking in the literature. Thus, we are not completely sure about the vacuum structure of the theory. In particular, we note that the answer to the following question has not been established: Is it enough to remove the unwanted vacua in Eqs. (\ref{eq:many-classical-vacua}) and (\ref{eq:many-classical-vacua-2}) in the four-dimensional theory? For additional degeneracy, due to the instantons that are already in the continuum, we do not have to remove such degeneracy on the lattice because it is physical. If any degeneracy of $\Phi_{\tt A} = 0$ that has no counterpart in the continuum other than the type of Eqs. (\ref{eq:many-classical-vacua}) and (\ref{eq:many-classical-vacua-2}), we should care about that.

In any event, if we introduce the supersymmetry breaking term Eq. (\ref{eq:Wilson-term}), the trivial vacuum is singled out, and we can proceed.

We can write down the $\cN = 2^*$ SYM action on the lattice in the following $Q^{(m)}$-exact form,
\bea
S_{\cN=2^*} &=& \beta_L \sum_\bn Q^{(m)} \Psi_L^{(m)}(\bn)~,
\eea
with $\beta_L$ denoting the lattice coupling and the lattice gauge fermion has the form
\bea
\Psi_L^{(m)}(\bn) &=&  \Tr \Big(\chi_{\tt A}(\bn) \Big[\hf \Phi_{\tt A}(\bn) - \qtr H_{\tt A}(\bn) - \frac{1}{8} \epsilon_{\tt ABC}[B_{\tt B}(\bn), B_{\tt C}(\bn)] - \frac{1}{4} [C(\bn), B_{\tt A}(\bn)] \Big] \nn \\
&& \quad \quad + \frac{1}{2\rtwo} \psi_\mu (D^{(+)}_\mu \phib)(\bn) - \frac{1}{4} \eta(\bn) [\phi(\bn), \phib(\bn)] + (\cV(\bn) + \cW(\bn) + \cY(\bn)) \nn \\
&& \quad \quad - \qtr \zeta(\bn) [C(\bn), \phib(\bn)] - \qtr \psi_{\tt A}(\bn) [B_{\tt A}(\bn), \phib(\bn)] + \cT(\bn) \nn \\
&&\quad \quad - \frac{1}{2 \rtwo} \chi_\mu(\bn) (D^{(+)}_\mu C(\bn)) + \frac{1}{2\rtwo} B_{\tt A}(\bn) (D \chi)_{\tt A}(\bn) \Big)~,
\eea
where
\bea
\cV(\bn) &=& -\qtr m \Big( (\psi_{\boldsymbol{3}}(\bn) - i \psi_{\boldsymbol{1}}(\bn)) (B_{\boldsymbol{3}}(\bn) + i B_{\boldsymbol{1}}(\bn)) \nn \\
&&~~~~~~~~~~ + (\psi_{\boldsymbol{2}}(\bn) - i \zeta(\bn))(B_{\boldsymbol{2}}(\bn) + i C(\bn)) \Big)~, \\
\cW(\bn) &=& \frac{i}{4} \Big( - \psi_{\boldsymbol{3}}(\bn) [\phib(\bn), B_{\boldsymbol{1}}(\bn)] + \psi_{\boldsymbol{1}}(\bn) [\phib(\bn), B_{\boldsymbol{3}}(\bn)] + \eta(\bn) [B_{\boldsymbol{3}}(\bn), B_{\boldsymbol{1}}(\bn)] \Big)~,~~~ \\
\cY(\bn) &=& \frac{i}{4} \Big( - \psi_{\boldsymbol{2}}(\bn) [\phib(\bn), C(\bn)] + \zeta(\bn) [\phib(\bn), B_{\boldsymbol{2}}(\bn)] + \eta(\bn) [B_{\boldsymbol{2}}(\bn), C(\bn)] \Big)~, \\
\cT(\bn) &=& \qtr \Big( (\chi_1(\bn) - i \chi_2(\bn))(H_1(\bn) + iH_2(\bn)) \nn \\
&&~~~~~~~~~~ + (\chi_3(\bn) + i \chi_4(\bn))(H_3(\bn) - iH_4(\bn)) \Big)~,
\eea
and 
\beq
(D \chi)_{\tt A}(\bn) \equiv 2 \Big( D^{(+)}_{\tt A} \chi_{\tt 4}(\bn) + \hf \epsilon_{\tt ABC} D^{(+)}_{\tt B} \chi_{\tt C}(\bn) \Big)~.
\eeq

It is straightforward to show that the lattice theory constructed here has no fermion doubling problem. The fermionic kinetic term of the theory is exactly the same as the one considered in Ref. \cite{Sugino:2003yb}, where the case of lattice $\cN=4$ SYM was discussed. There, it was shown that the fermion doubling problem does not occur in the lattice $\cN=4$ SYM theory based on Sugino discretization prescription.

We note that the lattice action of $\cN = 2^*$ SYM formulated here is gauge invariant, local, doubler free and exactly supersymmetric under one supersymmetry charge. However, the lattice theory is not twisted Lorentz invariant. Some of the mass terms of the theory contain twisted Lorentz indices that are uncontracted. The reason for twisted Lorentz symmetry breaking is the reduced R-symmetry of the $\cN=2^*$ SYM compared to that of the $\cN=4$ SYM. Although both theories are Lorentz invariant in their untwisted forms, one of them become twisted Lorentz noninvariant. We note that this does not lead to any inconsistency in the lattice formulation of $\cN=2^*$ SYM. The lattice theory is still Lorentz invariant. The continuum twisted theory is obtained by an exotic change of variables of the original Lorentz invariant theory. However, there are consequences for having twisted Lorentz symmetry breaking terms in the lattice theory. It will reduce the number of discrete symmetries of the lattice theory, and this in turn increase the number of unwanted operators that are allowed on the lattice. A careful listing of such operators and appropriate fine-tuning are needed before simulating the theory on the lattice. 
 
We also note that it would be possible to impose the admissibility condition \cite{Sugino:2004qd}
\beq
\label{eq:admissibility}
|| 1 - U_{\mu\nu} || < \epsilon
\eeq 
on each plaquette variable in order to solve the issues with vacuum degeneracy. We note that Eq. (\ref{eq:admissibility}) resolves the degeneracy (\ref{eq:many-classical-vacua}) and (\ref{eq:many-classical-vacua-2}) with keeping supersymmetry because the admissibility condition is imposed on the gauge fermion $\Psi^{(m)}$ of the $Q^{(m)}$-exact action and it does not affect the $Q^{(m)}$-exact structure. Reference \cite{Matsuura:2014pua} discusses another method to avoid the vacuum degeneracy while keeping supersymmetry.   

Although yet more vacua appear as discussed in Ref. \cite{Sugino:2004qd}, it is irrelevant to the discussion for the admissibility condition (\ref{eq:admissibility}).

We do not know which value should be chosen for $\epsilon$ in Eq. (\ref{eq:admissibility}) because we do not know the vacuum structure of $\Phi_A = 0$. The value of $\epsilon$ in the admissibility condition should be determined so as to exclude the unphysical vacua from Eqs. (\ref{eq:many-classical-vacua}) and (\ref{eq:many-classical-vacua-2}).

For the case that the gauge field has no topologically nontrivial structure (zero Pontryagin index), we think that it would be enough to remove the degeneracy (\ref{eq:many-classical-vacua}) and (\ref{eq:many-classical-vacua-2}) even in four dimensions and Eq. (\ref{eq:admissibility}) would be available since the nontriviality from the lattice version of the instantons could be irrelevant. In the numerical simulations, it would be a good starting point to try to simulate the lattice action with the boundary conditions of topologically trivial gauge fields and with the use of Eq. (\ref{eq:Wilson-term}) or the supersymmetry preserving Eq. (\ref{eq:admissibility}). 

%%%%%%%%%%%%%%%%%%%%%%%%%%%%%%%%%%%%%%%%%%%%%%%%%%%%%%%%%%%%%%%%%%%%%%
\section{Conclusions and Future Directions}
\label{sec:conclusions-future}
%%%%%%%%%%%%%%%%%%%%%%%%%%%%%%%%%%%%%%%%%%%%%%%%%%%%%%%%%%%%%%%%%%%%%%

In this work, we have provided a Euclidean lattice formulation of four-dimensional $\cN = 2^*$ SYM. The lattice formulation is gauge invariant, local, supersymmetric under one scalar supercharge and free from fermion doublers. We have also provided the continuum twisted formulation of $\cN = 2^*$ SYM starting from the Vafa-Witten twist of the $\cN = 4$ SYM theory. According to our knowledge, this is the first time such a continuum twisted formulation of $\cN=2^*$ SYM is presented. The lattice theory is obtained by transporting the twisted $\cN = 2^*$ SYM theory on to the lattice. The gauge field is placed on an oriented link, and all other fields are placed on the sites of the hypercubic unit cell. The covariant derivative operators are mapped to covariant difference operators on the lattice. The advantage of twisting is that we can preserve a part of the supersymmetry algebra, involving one of the scalar supercharges that results from twisting, on the lattice. 

We note that the lattice theory constructed here contains terms that are not twisted Lorentz invariant. We emphasize that this does not lead to any inconsistencies in the formulation. The twisted theory is still Euclidean Lorentz invariant since twisting is an exotic change of variables and the original untwisted $\cN=2^*$ SYM is Lorentz invariant. The presence of twisted Lorentz noninvariant terms is due to the reduced R-symmetry in the theory, which is $SU(2) \times U(1)$. There will be more counterterms generated on the lattice due to less symmetry, and one has to count the number of such terms and fine tune them before embarking on lattice simulations. A careful analysis about the amount of symmetries present in the lattice theory is still needed. The theory is invariant under eight supercharges in the continuum and on the lattice it is invariant under only one supercharge. A careful study is needed to see how these broken supercharges emerge as the continuum limit is taken. We save these investigations for future work. One should also check that the theory does not suffer from the sign problem.

At present there exist computer codes that can simulate four-dimensional $\cN=4$ SYM. However, this code is based on the B-model (geometric Langlands) twist of $\cN=4$ SYM and employs geometric discretization, in which fermions live on sites and oriented links of the unit cell of the lattice. The work presented here utilizes the A-model twist of $\cN=4$ SYM, and the discretization prescription is the one given by Sugino. According to our knowledge, a computer code for $\cN=4$ lattice SYM based on Sugino lattice action does not exist. However, there exist computer codes in one and two dimensions that can simulate maximally supersymmetric Yang-Mills based on the Sugino lattice prescription. See Refs. \cite{Kadoh:2015mka, Kadoh:2016eju, Kadoh:2017mcj, Kadoh:2017tgc} for some interesting physics results produced based on this code. 

The nonperturbative construction of four-dimensional $\cN = 2^*$ SYM discussed here can be used to simulate the theory at any finite value of the gauge coupling, mass parameter and number of colors. It would be interesting to simulate the lattice $\cN = 2^*$ SYM theory and study the observables related to the AdS/CFT correspondence.  

We note that there are many aspects of $\cN = 2^*$ SYM which would be interesting to study on the lattice. In Ref. \cite{Hoyos:2011uh}, it was discussed that the $N \to \infty$ theory, which has a holographic dual, evidently has no thermal phase transition at any nonzero temperature. But for finite values of $N$, there should be a distinct low temperature phase. Seeing evidence of this from lattice gauge theory simulations, and gaining information about the $N$ dependence of the transition would be interesting.

%%%%%%%%%%%%%%%%%%%%%%%%%%%%%%%%%%%%%%%%%%%%%%%%%%%%%%%%%%%%%%%%%%%%%%
\acknowledgments 

We are very thankful to Fumihiko Sugino for a careful reading of an earlier version of the manuscript and providing valuable feedback. We thank discussions with Simon Catterall, Poul Damgaard, Joel Giedt, Victor Giraldo, So Matsuura, David Schaich and Larry Yaffe. We gratefully acknowledge support from the International Centre for Theoretical Sciences (ICTS-TIFR), the Infosys Foundation and the Indo-French Centre for the Promotion of Advanced Research (IFCPAR/CEFIPRA).
%%%%%%%%%%%%%%%%%%%%%%%%%%%%%%%%%%%%%%%%%%%%%%%%%%%%%%%%%%%%%%%%%%%%%%

%%%%%%%%%%%%%%%%%%%%%%%%%%%%%%%%%%%%%%%%%%%%%%%%%%%%%%%%%%%%%%%%%%%%%%
\appendix  
%%%%%%%%%%%%%%%%%%%%%%%%%%%%%%%%%%%%%%%%%%%%%%%%%%%%%%%%%%%%%%%%%%%%%%

%%%%%%%%%%%%%%%%%%%%%%%%%%%%%%%%%%%%%%%%%%%%%%%%%%%%
\section{Gravitational Dual of $\cN=2^*$ SYM}
\label{sec:grav-dual}
%%%%%%%%%%%%%%%%%%%%%%%%%%%%%%%%%%%%%%%%%%%%%%%%%%%%

In this section we give a brief review of the gravitational dual of $\cN = 2^*$ SYM theory, at zero and finite temperatures.

%%%%%%%%%%%%%%%%%%%%%%%%%%%%%%%%%%%%%%%%%%%%%%%%%%%%
\subsection{Zero Temperature}
%%%%%%%%%%%%%%%%%%%%%%%%%%%%%%%%%%%%%%%%%%%%%%%%%%%%

The holographic dual of $\cN = 2^*$ SYM theory at zero temperature was constructed by Pilch and Warner \cite{Pilch:2000ue}. The dual geometry is a warped product of a deformed $AdS_5$ and a deformed 5-sphere. The deformed 5-sphere is foliated by elongated 3-spheres, whose $SU(2) \times U(1)$ isometry of which realizes geometrically the R-symmetry of the dual $\cN = 2^*$ SYM theory. The ``uplift'' of the five-dimensional supergravity to ten dimensions was also successfully constructed by Pilch and Warner. In the full ten-dimensional type IIB supergravity the two scalars are Kaluza-Klein modes, which deform the $AdS_5 \times S^5$ geometry dual to the $\cN = 4$ SYM theory. We can consider the dual theory as Einstein gravity coupled to two real supergravity scalars, which we denote as $\alpha$ and $\chi$, in five dimensions. The holographic dual of $\cN = 2^*$ SYM theory was well explored in Refs. \cite{Pilch:2000ue, Khavaev:1998fb, Buchel:2000cn, Evans:2000ct}.

We can also interpret the above-mentioned gravitational background as a dual description of $\cN = 4$ SYM theory perturbed by two relevant operators: a {\it bosonic} operator $\cO_2$ and a {\it fermionic} operator $\cO_3$. The supergravity scalars can be interpreted as bosonic and fermionic deformations of the D3-brane geometry. According to the general framework of holographic renormalization group flows \cite{Maldacena:1997re, Aharony:1999ti}, the asymptotic boundary behavior of scalars $\alpha$ and $\chi$ contains information about the couplings and expectation values of the dual operators $\cO_2$ and $\cO_3$ in the boundary gauge theory. 
  
The appropriate terms in the five-dimensional supergravity action, including the scalars $\alpha$ and $\chi$, can be written as
\beq
\label{eq:sugra-action}
I_5 = \frac{1}{4 \pi G_5} \int_{\cM_5} d \xi^5 \sqrt{-g} \left( \qtr R - \cL_{\rm matter} \right)~,
\eeq  
where the matter Lagrangian is
\beq
\cL_{\rm matter} = - 3 (\partial \alpha)^2 - (\partial \chi)^2 - \cP~,
\eeq 
with the potential
\beq
\cP = \widehat{g}^2 \left( \frac{1}{16} \left[ \frac{1}{3} \left( \frac{\partial W}{\partial \alpha} \right)^2 + \left( \frac{\partial W}{\partial \chi}\right)^2 \right] - \frac{1}{3} W^2 \right)
\eeq
determined by the superpotential
\beq
W = - e^{2 \alpha} - \hf e^{4 \alpha} \cosh(2 \chi)~.
\eeq

The dimensionful gauged supergravity coupling is 
\beq
\widehat{g}^2 = \left(\frac{2}{L}\right)^2~,
\eeq
where $L$ is the radius of the 5-sphere, and the 5-dimensional Newton's constant is 
\beq
G_5 \equiv \frac{G_{10}}{2^5~{\rm vol}_{S^5}} = \frac{\pi L^3}{2 N^2}~.
\eeq

From the action Eq. (\ref{eq:sugra-action}), we have the Einstein's equations
\beq
R_{\mu\nu} = 12 \partial_\mu \alpha \partial_\nu \alpha + 4 \partial_\mu \chi \partial_\nu \chi + \frac{4}{3} g_{\mu\nu} \cP~,
\eeq
and the equations for the scalars
\beq
\square \alpha = \frac{1}{6} \frac{\partial \cP}{\partial \alpha}~,~~\square \chi = \frac{1}{2} \frac{\partial \cP}{\partial \chi}~.
\eeq

%%%%%%%%%%%%%%%%%%%%%%%%%%%%%%%%%%%%%%%%%%%%%%%%%%%%
\subsection{Finite Temperature}
%%%%%%%%%%%%%%%%%%%%%%%%%%%%%%%%%%%%%%%%%%%%%%%%%%%%

The supergravity background geometry dual to finite temperature $\cN = 2^*$ SYM theory was constructed by Buchel and Liu in Ref. \cite{Buchel:2003ah}. When the temperature goes to zero, this geometry becomes the Pilch-Warner geometry \cite{Pilch:2000ue}. One can construct a map between finite temperature $\cN = 2^*$ SYM theory parameters and the parameters of the dual nonextremal geometry \cite{Buchel:2000cn, Buchel:2003ah}.

There are three supergravity parameters uniquely determining a nonsingular RG flow in the dual nonconformal gauge theory \cite{Buchel:2007vy}. They are unambiguously related to the three physical parameters of the $\cN = 2^*$ SYM theory: the temperature $T$, the bosonic mass $m_b$ and the fermionic mass $m_f$. (Note that for the case of $\cN = 2^*$ SYM theory we have strictly $m \equiv m_b = m_f$. It is still possible to consider the theory with unequal $m_b$ and $m_f$. The resulting theory will of course break supersymmetry further.)

In Ref. \cite{Buchel:2007vy} the thermal Pilch-Warner flow was investigated near the boundary of the supergravity geometry. From the asymptotic expansions near the boundary it is possible to identify the conformal weight-2 supergravity scalar, defined as $\alpha \equiv \log \rho$, as dual to turning on a mass for the bosonic components of the $\cN = 2^*$ hypermultiplet. The asymptotic expansion of $\rho$ contains parameters $\rho_{11}$ and $\rho_{10}$ \cite{Buchel:2007vy}, which can be interpreted as the coefficients of its non-normalizable and normalizable modes, respectively. The conformal weight-1 supergravity scalar $\chi$ can be identified as dual to turning on a mass for the fermionic components of the $\cN = 2^*$ hypermultiplet \cite{Buchel:2007vy}. 

Once the potential $\cP$ and the superpotential $W$ are given, it is possible to consistently truncate the finite temperature supergravity system to a purely bosonic deformation. This corresponds to the choice $\chi = 0$. However, it is inconsistent, beyond the linear approximation, to set the bosonic deformation to zero, that is, setting $\alpha = 0$ while keeping a fermionic deformation.

%%%%%%%%%%%%%%%%%%%%%%%%%%%%%%%%%%%%%%%%%%%%%%%%%%%%%%%%%%%%%%%
\subsection{Relating Supergravity and Gauge Theory Parameters}
%%%%%%%%%%%%%%%%%%%%%%%%%%%%%%%%%%%%%%%%%%%%%%%%%%%%%%%%%%%%%%%

The relation between $\cN = 2^*$ SYM theory and the supergravity parameters of the thermal Pilch-Warner geometry was established by Buchel {\it et al.} in Ref. \cite{Buchel:2000cn} and later by Buchel and Liu in Ref. \cite{Buchel:2003ah}.

Finite temperature softly breaks supersymmetry. Thus we could generalize the thermal $\cN = 2^*$ SYM theory by allowing different masses, $m_b$ and $m_f$, for the bosonic and fermionic components of the $\cN = 2^*$ hypermultiplet. Note that it is only when $m_b = m_f \equiv m$ and $T = 0$ that we have $\cN = 2$ supersymmetry. 

Turning on the bosonic and fermionic masses for the components of the $\cN = 2$ hypermultiplet sets a strong coupling scale $\Lambda$ in the theory. In this case, we could expect two qualitatively different thermal phases of the gauge theory. It depends on on whether $T \gg \Lambda$ or $T \ll \Lambda$. When $T \gg \Lambda$, we expect the thermodynamics to be qualitatively (and quantitatively when $T/\Lambda \to \infty$) similar to that of the $\cN = 4$ SYM theory plasma. When $T \sim \Lambda$ and $m_f = 0$, we expect an instability in the system. Turning on only the supergravity scalar $\alpha$, that is, setting $m_b \neq 0$ and $m_f = 0$, corresponds to giving positive mass squared to four $\cN = 4$ SYM scalars (the bosonic components of the $\cN = 2$ hypermultiplet). At the same time, the remaining two $\cN = 4$ SYM scalars acquire a negative mass squared. That is, they are the tachyons at zero temperature. However, at high enough temperatures, the thermal corrections would come into effect and stabilize these tachyons. As the temperature is lowered, we expect the reemergence of these tachyons. This is due to the fact that dynamical instabilities in thermal systems can manifest as thermodynamic instabilities. (See Ref. \cite{Gubser:2000ec} for arguments leading to this.) It was argued in Ref. \cite{Buchel:2005nt} that in general, thermodynamic instabilities are reflected to developing $c_s^2 < 0$, where $c_s$ is the speed of sound waves in the thermal gauge theory plasma. 
 
%%%%%%%%%%%%%%%%%%%%%%%%%%%%%%%%%%%%%%%%%%%%%%%%%%%%%%%%%%%%%%%%%%%%%%
\section{Euclidean Spinor Conventions and Mass Terms}
\label{sec:spinor-mass}
%%%%%%%%%%%%%%%%%%%%%%%%%%%%%%%%%%%%%%%%%%%%%%%%%%%%%%%%%%%%%%%%%%%%%%

Following the conventions given in Ref. \cite{McKeon:2001pm} we define the Euclidean Dirac spinors $\lambda, \lambdab$ using Weyl spinors $\lambda_{i\alpha}, \lambdab^{i\dot{\alpha}}$, with $i = 1, 2$ denoting the internal symmetry index and $\alpha, \dot{\alpha} = 1, 2$ denoting the spinor indices 

\beq
\lambda = \left( \begin{array}{c}
\lambda_{1\alpha} \\
\lambdab^{2\dot{\alpha}} \end{array} \right),~~\lambdab = \lambda^\dagger \gamma_0 = \left( \begin{array}{cc}
\lambda_2^{\phantom{2}\alpha} & \lambdab^1_{\phantom{1}\dot{\alpha}} \end{array} \right)~.
\eeq

We also have $\left(\lambda^i_{\phantom{1}\alpha}\right)^* = - \lambdab_{i \dot{\alpha}}$ and $\left(\lambdab^i_{\phantom{1}\dot{\alpha}}\right)^* = \lambda_{i \dot{\alpha}}$.

With these conventions it is straightforward to show that the fermion mass terms 
\beq
\Tr \Big(- m \lambda_1^{\phantom{1}\alpha} \lambda_{2\alpha} - m \lambdab^1_{\phantom{1}\dot{\alpha}} \lambdab^{2\dot{\alpha}}\Big) \nn
\eeq
are Hermitian.  

Upon using the relations between the twisted and untwisted fermionic field variables
\begin{align*}
 \lambda_1^{\phantom{1}1} &= \psi_{12} + \frac{i}{2\rtwo} \psi_{23}~, & \lambda_{21} &= \psi_{12} - \frac{i}{2\rtwo} \psi_{23}~, \\
 \lambda_1^{\phantom{1}2} &= \psi_{13} + \frac{i}{2\rtwo} \zeta~, & \lambda_{22} &= \psi_{13} - \frac{i}{2\rtwo} \zeta~, \\
 \lambdab_1^{\phantom{1}\dot{\alpha}} &= \chi_{2\dot{\alpha}}~, & \lambdab^2_{\phantom{1}\dot{\alpha}} &= - \frac{1}{2\rtwo}\chi_{1\dot{\alpha}}~,
\end{align*}
we have the mass term
\bea
- m ~\Tr \lambda_1^{\phantom{1}\alpha} \lambda_{2\alpha} &=& - m \Tr \Big(\lambda_1^{\phantom{1}1} \lambda_{21} + \lambda_1^{\phantom{1}2} \lambda_{22} \Big) \nn \\
&=& - m ~\Tr \Big((\psi_{12} + \frac{i}{2\rtwo} \psi_{23})(\psi_{12} - \frac{i}{2\rtwo} \psi_{23}) \nn \\
&& \quad \quad + (\psi_{13} + \frac{i}{2\rtwo} \zeta)(\psi_{13} - \frac{i}{2\rtwo} \zeta) \Big) \nn \\
&=& - \frac{im}{2\rtwo} ~\Tr \Big(- \psi_{12}\psi_{23} + \psi_{23}\psi_{12} - \psi_{13} \zeta + \zeta \psi_{13} \Big) \nn \\
\label{eq:fermion-mass-1}
&=& \frac{i}{\rtwo} m ~\Tr (\psi_{12}\psi_{23} + \psi_{13}\zeta)~. 
\eea

Let us look at the following mass term
\beq
- m ~\Tr \Big(\lambdab^1_{\phantom{1}\dot{\alpha}} \lambdab^{2\dot{\alpha}}\Big)~.
\eeq

Upon expanding the indices and using the conventions
\bea
\epsilon^{12} &=& \epsilon^{\dot{1}\dot{2}} = \epsilon_{21} = \epsilon_{\dot{2}\dot{1}} = +1~, \\ 
\epsilon^{21} &=& \epsilon^{\dot{2}\dot{1}} = \epsilon_{12} = \epsilon_{\dot{1}\dot{2}} = -1~,
\eea
we have
\bea
\label{eq:chi-mass-untw}
- m ~\Tr \Big(\lambdab^1_{\phantom{1}\dot{\alpha}} \lambdab^{2\dot{\alpha}}\Big) &=& -m ~\Tr \Big(\lambdab^1_{\phantom{1}\dot{1}} \lambdab^2_{\phantom{1}\dot{2}} - \lambdab^1_{\phantom{1}\dot{2}}\lambdab^2_{\phantom{1}\dot{1}}\Big)~.
\eea

Using the relations between the untwisted and twisted spinors in Eq. (\ref{eq:chi-mass-untw}), we have
\bea
- m ~\Tr \Big(\lambdab^1_{\phantom{1}\dot{\alpha}} \lambdab^{2\dot{\alpha}}\Big) &=& \frac{m}{2\rtwo} ~\Tr \Big(\chi_{2\dot{1}}\chi_{1\dot{2}} - \chi_{2\dot{2}}\chi_{1\dot{1}}\Big)~.
\eea

Upon using 
\bea
&& \chi^{\phantom{\mu}}_{i \dot{\alpha}} = \sigma^\mu_{i\dot{\alpha}}\chi^{\phantom{\mu}}_\mu~,
\eea
with the Euclidean convention $\sigma^\mu \equiv (\vec{\sigma},~i{\mathbb I})$, the fermion mass term becomes
\bea
- m ~\Tr \Big(\lambdab^1_{\phantom{1}\dot{\alpha}} \lambdab^{2\dot{\alpha}}\Big) &=& \frac{m}{2\rtwo} ~\Tr \Big(\chi_{2\dot{1}}\chi_{1\dot{2}} - \chi_{2\dot{2}}\chi_{1\dot{1}}\Big) \nn \\
&=& \frac{m}{2\rtwo} ~\Tr \Big((\chi_1 + i\chi_2)(\chi_1 - i\chi_2) - (-\chi_3 + i \chi_4)(\chi_3 + i \chi_4)\Big) \nn \\
&=& \frac{m}{2\rtwo} ~\Tr \Big( - 2i \chi_1 \chi_2 + 2i \chi_3 \chi_4 \Big) \nn \\
&=& - \frac{im}{\rtwo} ~\Tr (\chi_1 \chi_2 - \chi_3 \chi_4)~.
\eea

%%%%%%%%%%%%%%%%%%%%%%%%%%%%%%%%%%%%%%%%%%%%%%%%%%%%%%%%%%%%%
\section{Deriving the Mass Terms of Twisted $\cN = 2^*$ SYM}
\label{sec:ELN-terms}
%%%%%%%%%%%%%%%%%%%%%%%%%%%%%%%%%%%%%%%%%%%%%%%%%%%%%%%%%%%%%

The $Q^{(m)}$ variations on the linear combinations of the fields have the form
\bea
Q^{(m)} (\chi_1 + i \chi_2) &=& 2 (H_1 + i H_2)~, \nn \\
Q^{(m)} (\chi_3 + i \chi_4) &=& 2 (H_3 + i H_4)~, \nn \\
Q^{(m)} (H_1 + i H_2) &=& - \rtwo [\phi, (\chi_1 + i \chi_2)] + \rtwo m (\chi_1 + i \chi_2)~, \nn \\
Q^{(m)} (H_3 + i H_4) &=& - \rtwo [\phi, (\chi_3 + i \chi_4)] + \rtwo m (\chi_3 + i \chi_4)~, \\
Q^{(m)} (B_{12} + i B_{23}) &=& \rtwo (\psi_{12} + i \psi_{23})~, \nn \\
Q^{(m)} (B_{13} + i C) &=& \rtwo (\psi_{13} + i \zeta)~, \nn \\
Q^{(m)} (\psi_{12} + i \psi_{23}) &=& - 2 [\phi, (B_{12} + i B_{23})] + 2 m (B_{12} + i B_{23})~, \nn \\
Q^{(m)} (\psi_{13} + i \zeta) &=& - 2 [\phi, (B_{13} + i C)] + 2 m (B_{13} + iC)~. \nn 
\eea

The $Q^{(m)}$ variation of the quantity $P \equiv \Tr (\chi_1 - i \chi_2)(H_1 + i H_2)$ will contain the mass term $- (i m/\rtwo) \Tr (\chi_1 \chi_2)$.

We have
\bea
Q^{(m)} P &=& \Tr \Big[(\chi_1 - i \chi_2)(H_1 + i H_2)\Big] \nn \\
&=& \Tr \Big[\Big(2 (H_1 - i H_2)(H_1 + i H_2) + \rtwo (\chi_1 - i \chi_2) [\phi, (\chi_1 + i \chi_2)] \Big) \nn \\
&&\quad \quad \quad \quad - \rtwo m (\chi_1 - i \chi_2)(\chi_1 + i \chi_2) \Big] \nn \\
&=& \Tr \Big[2 (H_1^2 + H_2^2) + \rtwo (\chi_1 [\phi, \chi_1] + \chi_2 [\phi, \chi_2]) - 2 \rtwo i m \chi_1 \chi_2 \Big]~.
\eea

Similarly the linear combination $R \equiv \Tr (\chi_3 + i \chi_4)(H_3 - i H_4)$ will contain the mass term $(i m / \rtwo) \Tr (\chi_3 \chi_4)$.

We have
\bea
Q^{(m)} R &=& \Tr \Big[(\chi_3 + i \chi_4)(H_3 - i H_4)\Big] \nn \\
&=& \Tr \Big[\Big(2 (H_3 + i H_4)(H_3 - i H_4) + \rtwo (\chi_3 + i \chi_4) [\phi, (\chi_3 - i \chi_4)] \Big) \nn \\
&&\quad \quad \quad \quad - \rtwo m (\chi_3 + i \chi_4)(\chi_3 - i \chi_4) \Big] \nn \\
&=& \Tr \Big(2 (H_3^2 + H_4^2) + \rtwo (\chi_3 [\phi, \chi_3] + \chi_4 [\phi, \chi_4]) + 2 \rtwo i m \chi_3 \chi_4 \Big)~.
\eea

Defining $\cT \equiv \qtr(P + R)$ we have
\bea
Q^{(m)} \cT &=& \hf H_\mu^2 + \frac{1}{2\rtwo} \chi_\mu [\phi, \chi_\mu] - \frac{i}{\rtwo} m (\chi_1 \chi_2 - \chi_3 \chi_4)~.
\eea

This reproduces two of the terms in the $\cN = 4$ twisted SYM action and also two of the mass terms that appear in the $\cN = 2^*$ SYM action.

Let us now consider the $Q^{(m)}$ variation of the product of the terms $A \equiv (\psi_{12} - i \psi_{23})$ and $B \equiv (B_{12} + i B_{23})$. 

We have
\bea
Q^{(m)} \Tr (- AB) &=& \Tr \Big[ \Big(2 [\phi, (B_{12} - i B_{23})] - 2 m (B_{12} - i B_{23})\Big) (B_{12} + i B_{23}) \nn \\
&&\quad \quad \quad \quad + \rtwo (\psi_{12} - i \psi_{23})(\psi_{12} + i \psi_{23}) \Big] \nn \\
&=& \Tr \Big[\Big(2 [\phi, B_{12}] - 2i [\phi, B_{23}] \Big) (B_{12} + i B_{23}) \nn \\
&&\quad \quad \quad \quad - 2 m B_{12}^2 - 2 m B_{23}^2 + 2 \rtwo i \psi_{12}\psi_{23} \Big] \nn \\
&=& \Tr \Big(- 2 \phi [B_{12}, B_{12}] - 2 \phi [B_{23}, B_{23}] \nn \\
&&- 2 m B_{12}^2 - 2 m B_{23}^2 + 4i \phi [B_{12}, B_{23}] + 2 \rtwo i \psi_{12}\psi_{23} \Big)~.
\eea 

Let us compute the $Q^{(m)}$ variation of the product of the terms $D \equiv (\psi_{13} - i \zeta)$ and $E \equiv (B_{13} + i C)$.

We have
\bea
Q^{(m)} \Tr (-DE) &=& \Tr \Big[-\Big(-2 [\phi, (B_{13} - i C)] + 2 m (B_{13} - iC)\Big)(B_{13} + i C) \nn \\
&&\quad \quad \quad \quad -(\psi_{13} - i \zeta) \rtwo (\psi_{13} + i \zeta) \Big] \nn \\
&=& \Tr \Big(2 [\phi, B_{13}] B_{13} + 2i [\phi, B_{13}] C - 2i [\phi, C] B_{13} + 2 [\phi, C] C \nn \\
&&\quad \quad \quad \quad - 2m B_{13}^2 - 2m C^2 + 2 \rtwo i \psi_{13} \zeta \Big) \nn \\
&=& \Tr \Big(- 2 \phi [B_{13}, B_{13}] - 2 \phi [C, C] + 4i \phi [B_{13}, C] \nn \\
&&\quad \quad \quad \quad - 2 m B_{13}^2 - 2m C^2 + 2 \rtwo i \psi_{13} \zeta \Big)~.
\eea

Simplifying the terms
\bea
Q^{(m)} \Tr (-DE) &=& \Tr \Big(- 2 m B_{13}^2 - 2m C^2 - 2 \phi [B_{13}, B_{13}] - 2 \phi [C, C] \nn \\
&&\quad \quad \quad \quad + 4i \phi [B_{13}, C] + 2 \rtwo i \psi_{13} \zeta \Big)~.
\eea

Let us combine the terms
\bea
Q^{(m)} \Tr - (AB + DE) &=& \Tr \Big[- 2 m B_{\mu\nu}^2 - 2m C^2 - 2 \phi [B_{\mu\nu}, B_{\mu\nu}] - 2 \phi [C, C] \nn \\
&&+ 4i \phi \Big([B_{12}, B_{23}] + [B_{13}, C] \Big) + 2 \rtwo i (\psi_{12}\psi_{23} + \psi_{13} \zeta) \Big]~.~~~
\eea

Defining
\bea
\cV &\equiv& - \qtr m (AB + DE) \nn \\
&=& -\qtr m \Big( (\psi_{12} - i \psi_{23}) (B_{12} + i B_{23}) + (\psi_{13} - i \zeta)(B_{13} + i C)\Big)~,
\eea

we have

\bea
Q^{(m)} \Tr \cV &=& \Tr \Big(- \hf m^2 B_{\mu\nu}^2 - \hf m^2 C^2 \nn \\
&& - \hf m \phi [B_{\mu\nu}, B_{\mu\nu}] - \hf m \phi [C, C] \nn \\
&& + i m \phi \Big([B_{12}, B_{23}] + [B_{13}, C] \Big) \nn \\
&& + \frac{i}{\rtwo} m (\psi_{12}\psi_{23} + \psi_{13} \zeta) \Big)~.
\eea

Let us consider the $Q^{(m)}$ variation of the term
\bea
\cW &\equiv& \frac{i}{4} \Tr \Big( - \psi_{12} [\phib, B_{23}] + \psi_{23} [\phib, B_{12}] + \eta [B_{12}, B_{23}] \Big)~.
\eea

We have
\bea
Q^{(m)} \cW &=& \frac{i}{4} \Tr Q^{(m)} \Big( - \psi_{12} [\phib, B_{23}] + \psi_{23} [\phib, B_{12}] + \eta [B_{12}, B_{23}] \Big) \nn \\
&=& \frac{i}{4} \Big( 2 [\phi, B_{12}] [\phib, B_{23}] - 2m B_{12} [\phib, B_{23}] + \rtwo \psi_{12} [\eta, B_{23}] + \rtwo \psi_{12} [\phib, \psi_{23}] \nn \\
&& - 2 [\phi, B_{23}] [\phib, B_{12}] + 2 m B_{23} [\phib, B_{12}] - \rtwo \psi_{23} [\eta, B_{12}] - \rtwo \psi_{23} [\phib, \psi_{12}] \nn \\
&& - 2 [\phi, \phib] [B_{12}, B_{23}] - \rtwo \eta [\psi_{12}, B_{23}] - \rtwo \eta [B_{12}, \psi_{23}] \Big)~.
\eea

The terms involving fermions cancel among each other under the trace. Upon using the identity
\bea
\Tr \Big( [\phi, B_{12}] [\phib, B_{23}] - [\phi, B_{23}] [\phib, B_{12}] - [B_{12}, B_{23}] [\phi, \phib] \Big) = 0~,
\eea
we obtain
\bea
Q^{(m)} \cW &=& im \phib [B_{12}, B_{23}]~.
\eea

Similarly, the $Q^{(m)}$ variation of the term
\bea
\cY &\equiv& \frac{i}{4} \Tr \Big( - \psi_{13} [\phib, C] + \zeta [\phib, B_{13}] + \eta [B_{13}, C] \Big)~,
\eea
gives the mass term
\bea
Q^{(m)} \cY &=& im \phib [B_{13}, C]~.
\eea

%%%%%%%%%%%%%%%%%%%%%

%%%%%%%%%%%%%%%%%%%%%%

\end{document}